\documentclass[10pt]{article}
\input epsf

\begin{document}

\title{A simple model for hydromagnetic instabilities in the presence
of a constant magnetic field}
\author{A. Sandoval-Villalbazo$^a$, L.S. Garc\'{\i}a-Col\'{\i}n$^{b,\,c}$ \\
and A. Arrieta-Ostos $^a$ \\
$^a$ Departamento de F\'{\i}sica y Matem\'{a}ticas, Universidad Iberoamericana \\
Lomas de Santa Fe 01210 M\'{e}xico D.F., M\'{e}xico \\
E-Mail: alfredo.sandoval@uia.mx, anabel.arrieta@uia.mx\\
$^b$ Departamento de F\'{\i}sica, Universidad Aut\'{o}noma Metropolitana \\
M\'{e}xico D.F., 09340 M\'{e}xico \\
$^c$ El Colegio Nacional, Centro Hist\'{o}rico 06020 \\
M\'{e}xico D.F., M\'{e}xico \\
E-Mail: lgcs@xanum.uam.mx} \maketitle
\bigskip
\bigskip

\begin{abstract}
In this paper we study a simple model consisting of a dilute fully
ionized plasma in the presence of the gravitational and a constant
magnetic field to analyze the propagation of hydromagnetic
instabilities. In particular we show that the so called Jeans
instability is in principle affected by the presence of the
magnetic field. A brief discussion is made attempting to assess
this influence in the stage of the evolution of the Universe where
structures were formed. The most logical conclusion is that if
magnetic fields existed in those times their magnitudes were too
small to modify Jeans' mass. Our results places limits of the
possible values of seed magnetic fields consistent with the
formation structures in the Universe. These values are within the
range of the results obtained by other authors.
\end{abstract}

\section{Introduction}
\label{sec:intro}

Magnetic fields have a significant effect on virtually all astrophysical
objects. They are observed in all scales. Close to home, the Earth has a
bipolar magnetic field with a strength of $0.3 G$ at the equator and
$0.6 G$ at the poles (Carilli \& Taylor 2002). Within the interstellar
medium, magnetic fields are thought to regulate star formation via the
ambipolar diffusion mechanism (Spitzer 1978). Our Galaxy has a typical
interstellar magnetic field strength of $\sim 2 \mu G$ in both regular
ordered and random components. Other spiral galaxies have been estimated
to have magnetic field strengths of $5$ to $10 \mu G$, with fields
strengths up to $50 \mu G$ found in starburst galaxy nuclei
(Beck et al. 1996). Also magnetic fields are fundamental to the observed
properties of jets and lobes in radio galaxies, and they may be primary
elements in the generation of relativistic outflows from accreting massive
black holes (Carilli \& Taylor 2002).

Magnetic fields with typical strength of order 1$\mu G$  have been
measured in the intercluster medium using a variety of techniques.
Large variations in the field strength and topology are expected from
cluster to cluster, especially when comparing dynamically relaxed
clusters to those that have recently undergone a merger.  Magnetic fields
with strengths of $10-40 \mu G$ have been observed in some
locations (Carilli \& Taylor 2002). In all cases, the magnetic fields
play important role in the energy transport in the intercluster medium
and in gas collapse.

On the other hand at the cosmological level the presence or existence
of magnetic fields is more controversial. In a recent review on the subject
(Widrow, 2002) it is firmly asserted that a true cosmological magnetic field
is one that cannot be associated with collapsing or virialized structures.
Thus the particular role that they may play in the epoch of galaxy formation
is rather obscure. Although limits have been placed on the strength of
cosmological magnetics fields from Faraday rotation studies, of high redshift
sources, anisotropy measurements of the CMB and the light element
abundances from nucleosynthesis, the question remains:
Is there the possibility that the Jeans mass arising from gravitational
instabilities responsible for galaxies formation be modified by the
presence of a magnetic field?.

In spite of the dubious background provided by our present
knowledge, this question has been tackled since over fifty years
ago. In fact, already Chandrasekhar \& Fermi (1953) reached the
conclusion that Jeans criteria for the onset of a hydrodynamic
instability is unaffected by a magnetic field in an extended
homogeneous gas of infinite conductivity in the presence of an
uniform magnetic field. However, in their calculation they assumed
that within the gas there existed a fluctuating magnetic field.
This problem has been retaken by several other authors in
different contexts. In particular Lou (1996) studied the problem
of gravitational collapse in a magnetized dynamic plasma in the
presence of a finite amplitude circular polarized Alfv\'en wave.
This author does find a case in which Jeans wave number $k_J$ is
indeed modified by the magnetic field by a term proportional to
$[c_0^2 + c_A^2]^{1/2}$ where $c_0$ is the velocity of sound and
$c_A= B z_0 (4 \pi \rho_0)^{-1/2}$ is Alfv\'en wave speed $Bz_0$
being the z-component of the uniform magnetic field. Other
attempts to show that magnetic fields do play an essential
ingredient in galaxy formation have been performed, Kim, Olinto \&
Rosner (1996), although not specifically addressing the question
of a magnetic instability. Tsagas and Maartens (2000) have
performed a magnetohydrodynamical  analysis within a relativistic
framework addressing Jeans instability on the basis of previous
work, Tsagas and Barrow (1997,1998).

In view of all these efforts we still feel that the simple question
of whether or not a dilute non-magnetized plasma cloud placed in the
presence of an external, uniform magnetic field in which density
fluctuations are also present due to a fluctuating gravitational field,
exhibits a Jeans wave number which is modified by the presence of the field,
has not yet been fully discussed in the literature. This is the purpose
of the present work. The basic and rather simple formalism is given in
\S~\ref{sec:formal}. \S~\ref{sec:disp} is devoted to the derivation
of the dispersion relation leading to the modified form of $k_J$ and some
attempts to place the relevance of the results within a realistic frame
for existing magnetic field intensities. Some concluding remarks are given
in \S~\ref{sec:conc}.

\section{Basic Formalism}
\label{sec:formal}

We start by assuming that the dynamics of the dilute plasma is governed
by Euler's equations of hydrodynamics, namely the balance equations for the
fluid's mass density $\rho (\vec{r},t)$ and its velocity $u(\vec{r},t)$.
Thus,

\begin{equation}
\frac{\partial \rho }{\partial t}+\nabla \cdot \left( \rho \vec{u}\right) =0
\label{eq:one}
\end{equation}

\begin{equation}
\frac{\partial \left( \rho \vec{u}\right) }{\partial t}+\nabla
\cdot \left( \rho \vec{u}\vec{u}\right)+\nabla p
=\vec{f}_{g}+\vec{f }_{M}
\label{eq:two}
\end{equation}

In equation \ref{eq:two}, $\vec{f}_g $ is the force arising from
the gravitational field and we assume that the plasma is diluted enough
so that due to the enormous mass difference between the ions and the electrons,
the effect of the external field $\vec{B}$ will be substantially larger
on the former. Thus the Lorentz force $\vec{f}_M = {q \over m} \rho_0 (\vec{u}
\times \vec{B})$ where $m$ is the mass of the ions having charge q.

Eqs. (\ref{eq:one}-\ref{eq:two}) can be linearized by introducing density
and velocity fluctuations defined by:

\begin{equation}
\rho =\rho _{0}+\delta \rho
\label{eq:three}
\end{equation}

\begin{equation}
\vec{u}=\vec{u}_{0}+\delta \vec{u}
\label{eq:four}
\end{equation}

\noindent
and,

\begin{equation}
\delta \theta \equiv \nabla \cdot \left( \delta \vec{u}\right)
\label{eq:five}
\end{equation}

\noindent
where $\rho _{0}$ is the average density. The fluid is assumed to
be static, so that $\vec{u}_{0}=0$, $\varphi $
represents the gravitational potential and the external magnetic
force is that corresponding to a constant magnetic field
$\vec{B}=\left( B_{0}+\delta B\right) \hat{k}$, so that the
linearized equations for the density and velocity fluctuations can
be written as:

\begin{equation}
\frac{\partial \left( \delta \rho \right) }{\partial t}+\rho _{0}\delta
\theta =0  \label{eq:six}
\end{equation}

\noindent
and

\begin{equation}
\rho _{0} \frac{\partial \left( \delta \vec{u}\right) }{\partial
t} +  \nabla \left( \delta p\right) =-\rho
_{0}\nabla \left( \delta \varphi \right) \nonumber\\
   + \frac{q}{m}\rho_{0}\left( \delta \vec{u}\times \vec{B}_{0}\right)
\label{eq:seven}
\end{equation}

\noindent where $\vec{f}_g=-\nabla ( \delta \varphi))$, $\delta
\varphi$ being the fluctuating gravitational field. Neglecting
temperature fluctuations, the pressure term in equation
(\ref{eq:seven}) may be rewritten in terms of the density
fluctuations through the local equilibrium assumption namely,
$p=p(\rho)$ so that

$$
\nabla p = \left({\partial p \over \partial \rho_0}\right)_T \nabla \rho =
{c_0^2 \over \gamma} \nabla \rho .
$$

\noindent We know recall that $K_T$, the thermal compressibility
satisfies the relation $K_T = \gamma / c_0^2$ where $\gamma =
C_p/C_v$ and $c_0$ is the velocity of sound in the medium.
Therefore, equation (\ref{eq:seven}) may be rewritten as

\begin{equation}
\rho_0 \frac{\partial \left( \delta \vec{u}\right) }{\partial t}
+\frac{c_0^2}{\gamma}\nabla \left( \delta \rho \right) =  - \rho_0
\nabla \left( \delta \varphi \right) +  \frac{q}{m} \rho_0 \left(
\delta \vec{u}\times \vec{B}_{0}\right). \label{eq:eight}
\end{equation}

\noindent Assuming now that $\delta \varphi $ is defined through
Poisson's equation so that $\nabla ^2 (\delta \varphi)=4\pi G
\delta \rho$ where G is the gravitational constant, that
$\vec{B}_o = B_o \hat k $ where $\hat k$ is the unit vector along
the z-axis and noticing that for this case the last term equals
$B_o(\hat i \delta u_y - \hat j \delta  u_x)$, equation
(\ref{eq:eight}) reduces to

\begin{equation}
-\rho_0 \frac{\partial \left( \delta \theta \right) }{\partial t}
+  \frac{c_0^2}{\gamma }\nabla^2 \left(\delta \rho\right) =  -
4\pi G\rho _0 \left(\delta \rho\right)  +  \frac{q}{m}B_0 \rho
_{0}\left( \nabla \times \delta \vec{u}\right)_ {\hat k }
\label{eq:nine}
\end{equation}

\noindent after taking its divergence and using equation
(\ref{eq:six}).

Equations (\ref{eq:six}) and (\ref{eq:nine}) are now two
simultaneous equations for $\delta \rho$ and $\delta \vec{u}$
which need to be solved. To do so we notice first that
$\left(\nabla \times \delta \vec{u}\right)_{\hat k }= -\hat k
\left[ \nabla(\delta \vec{u}) + {\partial (\delta u_z) \over
\partial z}\right]$ so that using eq. (\ref{eq:six}) we may write
that

\begin{equation}
 \frac{q}{m}B_0 \rho _{0}{ \partial \over \partial t}
 \left( \nabla \times \delta \vec{u} \right)_{\hat k } =
 -  \frac{q}{m}B_0 \rho _{0}\left[{\partial \over \partial t} \left( -{1\over \rho_0}
{\partial \rho \over \partial t}\right) + {\partial \over \partial t} {\partial
( \delta u_z ) \over \partial z} \right]
\label{eq:ten}
\end{equation}

Finally, taking the time derivate of eq. (\ref{eq:nine}) and using
eqs. (\ref{eq:six}) and (\ref{eq:ten})  one is led to the result
that

\begin{eqnarray}
- {\partial^3 \over \partial t^3}(\delta \rho) + {c_0^2 \over
\gamma } \nabla^2 \left( {\partial (\delta \rho) \over \partial t}
\right) + 4\pi G \rho_0 {\partial \over \partial t} (\delta \rho)
- \nonumber\\ \left( {q B_0 \over m} \right)^2 {\partial \over
\partial t} \left( {\partial(\delta \rho) \over \partial t
}\right) - \left( {q B_0  \over m} \right)^2 \rho_0 {\partial
\over \partial t} \left( {\partial(\delta u_z) \over \partial z
}\right) =0
\end{eqnarray}

Integrating once with respect to time and setting the integration constant
equal to zero which does not affect the validity to our argument, we get that

\begin{equation}
- {\partial^2 \over \partial t^2}(\delta \rho)
+ {c_0^2 \over \gamma } \nabla^2 (\delta \rho)\nonumber\\
+ 4\pi G \rho_0  (\delta \rho) - \left( {q B_0 \over m} \right)^2
\left( {\partial(\delta \rho) \over \partial t  }\right) - \left(
{q B_0 \over m} \right)^2 \rho_0  \left( {\partial(\delta u_z)
\over \partial z  }\right) =0 \label{eq:eleven}
\end{equation}

Equation (\ref{eq:eleven}) is now a single equation for the
density fluctuations $\delta \rho$. Indeed, since $B_0$ points
along the z-axis, $[\delta \vec{u} \times \vec{B}_0]_{\hat k} =0$
so that eq. (\ref{eq:eight}) reduces to,

\begin{equation}
\rho_0 {\partial (\delta u_z) \over \partial t} +
\left({c_0^2 \over \gamma} - 4\pi G \rho_0 \right){\partial(\delta \rho) \over \partial z}
=0
\label{eq:twelve}
\end{equation}

The solution to Eqs. (\ref{eq:eleven}) and (\ref{eq:twelve}) is
readily achieved proposing that $\delta \rho $ is described by a
plane wave namely

\begin{equation}
\delta \rho = A e^{i(\vec{k} . \vec{r} - \omega t )}
\label{eq:trece}
\end{equation}

Taking the partial derivative with respect to $z$ in eq.
(\ref{eq:twelve}), exchanging the time and space derivatives in
the first term and calculating ${\partial^2 \over \partial
z^2}(\delta \rho)$ we get that

\begin{equation}
\rho_0 {\partial \over \partial t} \left( {\partial \over \partial z}
(\delta u_z) \right)
+  \left({c_0^2 \over \gamma} -4\pi G \rho_0 \right)
\left( -k^2_z \delta \rho \right) =0 \nonumber
\end{equation}

This result implies that

\begin{equation}
\rho_0 {\partial \over \partial t} \left( {\partial \over \partial z}
(\delta u_z) \right) = k_z^2 \Gamma A e^{i(\vec{k} . \vec{r} - \omega t )} \nonumber
\end{equation}

\noindent
were $\Gamma = {c_0^2 \over \gamma }- 4\pi G \rho_0 $.

But the left hand in this expression must be real so that right
hand side must be proportional to $cos (\omega t)$. Thus, on the
average, for large times $cos (\omega t) =0$ so that to a first
approximation we can set ${\partial \over \partial z} (\delta u_z)
\sim 0$. With this approximation, Eqs. (\ref{eq:eleven}) and
(\ref{eq:trece}) yield the dispersion relation

\begin{equation}
\omega^{2}-{c_{0}^{2} \over \gamma} k^2 + 4 \pi G \rho_{0}
-\left(\frac{q B_0}{m}\right)^{2}=0 .
\label{eq:catorce}
\end{equation}

Instabilities enhanced by the gravitational and magnetic field arise when
the roots for $\omega $ in this equation are imaginary. The threshold
value of $k$ beyond which this happens is precisely Jeans wave
number and is here given by

\begin{equation}
k_{J}^{2}=\frac{\gamma}{c_0^2}\left[4 \pi G \rho_{0}
-\left(\frac{qB_0}{m}\right)^{2}\right] .
\label{eq:quince}
\end{equation}

\noindent
Clearly, if $B_0 = 0$ we recover the well known expression for $k_J$.
We wish to stress here that the approximation taken above does not
imply that there are no velocity fluctuations along the $z$ axis,
the direction along which the magnetic field is acting, only that
their gradient along such a direction is negligible. Due to the results
to be discussed hereafter, we believe that a more detailed analysis
withdrawing this assumption is not necessary.
The question now is how relevant is the second term in hindering
structure formation. This will be analyzed in the following section.

\section{Analysis of the Dispersion Relation}
\label{sec:disp}

As indicated in the previous section, the linearized version of
fluctuating non-dissipative hydrodynamics predicts that for a
dilute non-magnetized homogeneous fully ionized plasma in the
presence of a uniform magnetic field leads to a Jeans wave number
which is, as depicted in equation (\ref{eq:quince}), a competition
between the gravitational and magnetic fields. The question is if
this result has any bearing on the value of Jeans mass in
realistic cases. Clearly equation (\ref{eq:quince}) points at two
possibilities, namely, if the magnetic term is  negligible  or of
the same order as the gravitational one. As we recall, Jeans mass
is defined as

$$ M_J \equiv {4 \pi \over 3} \rho_o \lambda_J^3= {4 \pi \over 3} \rho_o \left(
{2 \pi \over k_J} \right)^3$$

\noindent so that using equation (\ref{eq:quince})

\begin{equation}
M_J = {32 \pi^4 \over 3} \rho_o \left[{c_o^2 \over \gamma}{1 \over
4 \pi G \rho_o - \left({qB_o \over m}\right)^2}\right]^{3/2}
\label{eq:dseis}
\end{equation}

\noindent
where $\rho_o=mn_o$, $n_o$ being the particle density in the plasma.
As it has been exhaustively discussed in the literature (Jeans 1945,
Sandoval-Villalbazo \& Garc\'{\i}a-Col\'{\i}n 2002) without the magnetic
contribution, present values of $\rho_o \sim 10^{-29} gr/cm^3$, $m=m_H
\sim 10^{-24} gr$ and $T\sim 10^5 K$ yield for $M_J \sim 10^{11} M_\odot$.

Nevertheless, a close examination of eqs. (\ref{eq:quince}) and
(\ref{eq:dseis}) requires some thought. Firstly, one should notice
that in order for these results be physically meaningful, $4\pi
\rho_0 G > \left( {q B_0 \over m}\right)^2 $ must be fulfilled.
This inequality involves two critical parameters namely, $\rho_0 $
and $B_0$ precisely at the stage where structures are beginning to
develop in the evolution of the Universe. Next, since the proton
charge-mass ratio is approximately equal to $10^8$ c/kg, even for
small fields the cyclotron frequency is quite large, unless $B_0$
is very small, of the order of $10^{-24}$G. Thus, we may think of
these results as putting a limit on the value of the seed fields
that could have existed when structures began to form. In fact,
since it is known (Silk 1980) that the average matter density
prevailing when our own galaxy was formed was approximately equal
to $10^{-22}$ km/m$^3$, Eqs. (\ref{eq:quince} - \ref{eq:dseis})
hold only if $B_0$ is of the order of $10^{-24}$ G. This is in
reasonable agreement with the conclusions reached by several
authors that have examined the connection between the creation of
the first fields and the formation of large scale structures.
Although the values reported seem to depend on the cosmological
model Widrow (2002) the several estimates seem to lie in the
interval $10^{30} < B_0 < 10^{-19} $G. According to our results if
$\rho_0 $ is around the value quoted above then, if magnetic
fields existed in the cosmos they could hardly exceed to value of
$10^{-24}$G if we take as $M_J \sim 10^{12}$ M$\odot$. If no seed
magnetic field existed then the standard and well established
result for Jeans mass obviously remains unaffected. Notice
however, that this parameter could be enhanced by seed magnetic
fields for which the denominator in Eq. (\ref{eq:dseis}) becomes
small but positive. This remains to be tested.

\section{Conclusions}
\label{sec:conc}

The simple model here discussed for a fully ionized dilute plasma
in the presence of a homogeneous uniform magnetic field
shows clearly the rather peculiar behavior of Jeans mass due
to the confining effect of the magnetic field. Although such effect could
be important this would require very high densities and small magnetic
fields. The densities required would be so high that the model itself
becomes dubious and such values are completely at odds with observations.
On the other hand for low densities the conclusion is that gravitational instabilities will occur in the absence of magnetic
fields or, as envisaged by some authors (Widrow 2002) possible cosmological
fields $\sim 10^{-24}$G. Yet, such fields have not yet been detected.

This work was partially supported by CONACyT (M\'{e}xico), project
41081-F.


\begin{thebibliography}{10}

\bibitem{} Beck, R., Brandenburg, A., Moss, D., Shukurov, A. \&
Sokoloff, D. 1996, Annu. Rev. Astron. Astrophys., 34, 155

\bibitem{} Carilli, C.L. \& Taylor, G. B. 2002, Annu. Rev. Astron.
Astrophys. 40, 319

\bibitem{} Chandrasekhar, S. \& Fermi, E. 1953,  ApJ, 118, 116

\bibitem{} de Groot, S.R. \& Mazur, P. Non-equilibrium thermodynamics, Dover Publications,
Mineola, N.Y. 1984

\bibitem{} Jeans J. 1945, Astronomy and Cosmogony, Dover Publications, Inc. p.645

\bibitem{} Kim, E., Olinto, A. \& Rosner, R. 1996, ApJ, 468, 28

\bibitem{} Lou, Y.-Q. 1996, MNRAS, 279, L67

\bibitem{} Peebles, P.J.E. 1993, Principles of Physical Cosmology, Princeton Univ. Press,
Princeton, N.J. 2nd. Edition

\bibitem{} Sandoval-Villalbazo, A. \& Garc\'{\i}a-Col\'{\i}n, L. S. 2002, Class. and
Quant. Gravity, 19, 2171

\bibitem{} Spitzer, L. 1978, Physical Processes in the Interstellar Medium.
New York: Wiley

\bibitem{} Silk J. 1980, ``The Big Bang''. W.H. Freeman and Co. San
Francisco

\bibitem{} Tsagas C. and  Barrow J.D. Class.Quant.Grav. 14 (1997)
2539-2562 gr-qc/9704015.

\bibitem{} Tsagas C. and  Barrow J.D. Class.Quant.Grav. 15 (1998)
3523-3544  gr-qc/9803032.

\bibitem{} Tsagas C. and Maartens R., Phys.Rev. 2000 D61  83519. astro-ph/9904390

\bibitem{} Weinberg, S. (1971), ApJ, 168, 175

\bibitem{} Widrow, L.M. 2002, Rev.Mod.Phys., 74,775

\end{thebibliography}
\end{document}